# Separation of Hoke and Schottky effects for improvement of mixed halide perovskite solar cell stability

V. Ivanov[1], E. Ageev[1], D. Danilov[2], E.Yu. Danilovsky[1], D.S. Gets[1]*

[1] School of Physics and Engineering, ITMO University, 49 Kronverksky pr., 197101, St. Petersburg, Russia

[2] Interdisciplinary Resource Centre for Nanotechnology, St. Petersburg State University, 7/9 Universitetskaya Nab., 199034 St. Petersburg, Russia;

E-mail: dmitry.gets@metalab.ifmo.ru

## Abstract

Ion migration in halide perovskites is a key factor limiting the operational stability of solar cells due to formation of halogen ion enriched domains and accumulation layers. The present work demonstrates the manifestation of ion migration in two ways via Hoke and Schottky effects. Both effects are induced by external exposure but have its peculiar way of solar cell performance degradation. We demonstrate the effects of ion migration on the device performance by measuring time dependent short-circuit current and different impedance characteristics that allow to see how charge-carrier separation property degrades. The Schottky effect leads to the rapid decrease of charge-carrier separation characteristic of solar cell while Hoke effect leads to the slow defect accumulation in the perovskite layer leading to the enhanced Shockley-Read-Hall recombination. Separation of these two effects can be realized by simple increase of transport layer thickness. A thick transport layer blocks losses of charge-carrier selectivity in a solar cell and leads to an enormous increase of $T_{80}$ time, from 15 seconds up to 60 minutes.

Introduction

Halide perovskite solar cells drastically changed the world of optoelectronics and for more than 10 years perovskite solar cells reached power conversion efficiencies (PCEs) of well-established technologies based on Si (silicon heterojunction, SHJ) or GaAs[1]. A lot of effort was put into the iodine-based perovskites like MAPbI$_3$ due to its optimal band gap of 1.6 eV allowing to achieve almost the highest efficiency according to Schokley-Quisser limit[2], which further allowed to establish fabrication of tandem solar cells with perovskite wide band gap top cell and bottom low band gap cell based on Si (1.12 eV) or CIGS (1-1.7 eV)[3]. Combination with iodine perovskite worked well for efficient utilization of infrared part of the Sun spectrum, as for the visible part the band gap of perovskite absorber must be wider. The proper values are around 2 eV, which corresponds to the mixed halide perovskites like MAPbBr$_2$I[4]. Unfortunately, mixed halide perovskites have several undesirable properties like strong ion migration and halide segregation (Hoke effect) that limits its applicability.

The halide segregation first discovered and described by Hoke[5] is a reversible process when perovskite anion ions (iodine and bromine) start to move under external excitation (applied electrical bias or light exposure)[6] and form domains enriched with one type of anion[7]. The formation of anion-enriched domains results in temporary changes of optical and electrical properties of perovskite material and therefore optoelectronic devices. After the excitation removal the domains disappear, and ions redistribute in the host perovskite. The Hoke effect dramatically affects the performance mixed halide perovskite optoelectronic devices since the anion-enriched domains, especially iodine ones possessing low band gap by driving the optical properties of segregated perovskite. Usually in solar cells the Hoke effect results in the drop of PCE[8] since the optical properties are driven by segregated regions[9].

The discovery of the Hoke effect led to intensive research of methods of how to withstand it[10]. The basic solution to cure the Hoke effect is the use of multication approach for development of mixed halide perovskites. The MA$^+$-cation offers the best tolerance factor and thus maintains perovskite phase stabilization, but it is considered as the most volatile one demonstrating the highest instability against light exposure leading to the immediate halide segregation. It is an intrinsic process of mixed halide perovskites with high bromine concentration ( > 50%) that only enhances by additional strain induced by anion-rich domain formation. The introduction of FA$^+$ cation and fabrication of double cation mixed halide perovskites results in improved crystalline structure[11], solar cell performance and significantly suspends the halide segregation, but the formation of multiple crystalline phases still promotes the halide segregation. The addition of small percentage of Cs and use of triple cation perovskites allow to regulate phase formation, stabilize the crystalline lattice and suppress halide segregation[12]. Despite all benefits of multication approach it still does not allow to fabricate stable mixed halide perovskites with higher band gap values such as 2 eV[13], which are of highest interest for development of 3 junction tandems solar cells[14].

The origin of Hoke effect lies in the cause of ion migration[9] and perovskite capacity to form anion-enriched domains[15]. Under light excitation the excess of charge carriers forms in the bulk of mixed halide perovskite film leading to the induction of an electrical field[9]. The anion ions, especially iodine, which have low activation energies of 0.33-0.58 eV[16], start to move compensating induced electrical field and producing additional point defects that enhance defect assisted recombination. In the case of the film without electrical contacts the ion movement has an omnidirectional way, while in the case of optoelectronic device due to the presence of the electrical bias produced by the contact potential difference the ion movement has designated direction towards electrodes according to the ion charge. In both cases anion enriched domains formed in the whole film. The movement of ions towards the electrodes results in various effects like passivation of defects at perovskite layer interfaces or ion accumulation[17]. The presence of charged ion layer at the perovskite layer can lead to the Schottky effect with the following modification of the device band structure[18]. The formation of ion accumulation layers lead to the strong modification of barrier property affecting its charge carrier separation property, which is an ability to block the movement of charge carriers to opposite electrode[19,20].

Generally, the performance of perovskite solar cells is evaluated by several main parameters: PCE, absence of hysteresis of current-voltage characteristic and PCE sustainability under sun exposure[21]. PCE is an obvious parameter that reflects the overall capability of device to convert absorbed light into electricity. The hysteresis (H-index) reflects how migrating ions change the PCE due to *in situ* perovskite passivation by ions[22]. PCE sustainability under Sun illumination shows how fast the device loses its efficiency during working conditions. But all these parameters give a general description neglecting the microscale manifestation of ion migration which has a lot of different outcomes on the device performance. It is argued that moving ions can change electronic structure of the device by passivation of surface defects[23] or formation of charged layers at perovskite interfaces[24]. It also leads to the formation of anion domains that affect the charge carrier recombination[25] and introduce additional defects at the domain interfaces[26] due to different lattice constants of host perovskite and iodine- or bromine-rich domains[27–29].

In this work we demonstrate the interplay of Hoke and Schottky effects originated from ion migration. We indicate how these effects can be separated by simple adjustments of device architecture and show the net value of each effect on the device's performance. The demonstration of these effects is conducted on the most segregation unstable perovskite MAPbBr$_2$I and the management of these effects can improve the device stability over 200 times. We show how simple enhancement of charge carrier separation property allows to improve the device stable performance from 15 seconds up to 1 hour.

Results and discussion.

For the demonstration of Hoke and Schottky effects manifestation on the performance of perovskite solar cells (SC) we consider MAPbBr$_2$I perovskite (Figure S1) due to its remarkable ion migration property[6] under sun illumination or applied electrical voltage. This results in Hoke effect and high instability of SC performance. Ion migration affects the device performance in two possible ways: defect accumulation[25] and modification of device band structure[30,31] (Figure 1). This can also lead to irreversible physical damage while measurements, in which high energy irradiation interacts with the material (Figure S2), although it is not the only reason[32].

Defects occur due to lattice distortion induced by halide enriched domain formation inside the host perovskite material, the Hoke effect[5]. The second one arises from the formation of pin-structure inside the perovskite layer due to ion migration towards the interfaces of perovskite layer with transport layers (TLs)[17]. The formed pin structure strongly affects the functionality of whole SC due to alternation of charge carrier separation property by formation of accumulation layers of charged ions at the perovskite interfaces, the Schottky effect. As a result, in both cases device demonstrate performance degradation, but Hoke effect should result in slow degradation of SC parameters upon sun illumination, while the Schottky effect manifests as rapid decrease of device performance due to negative influence on the basic working principle of SC, charge carrier separation.

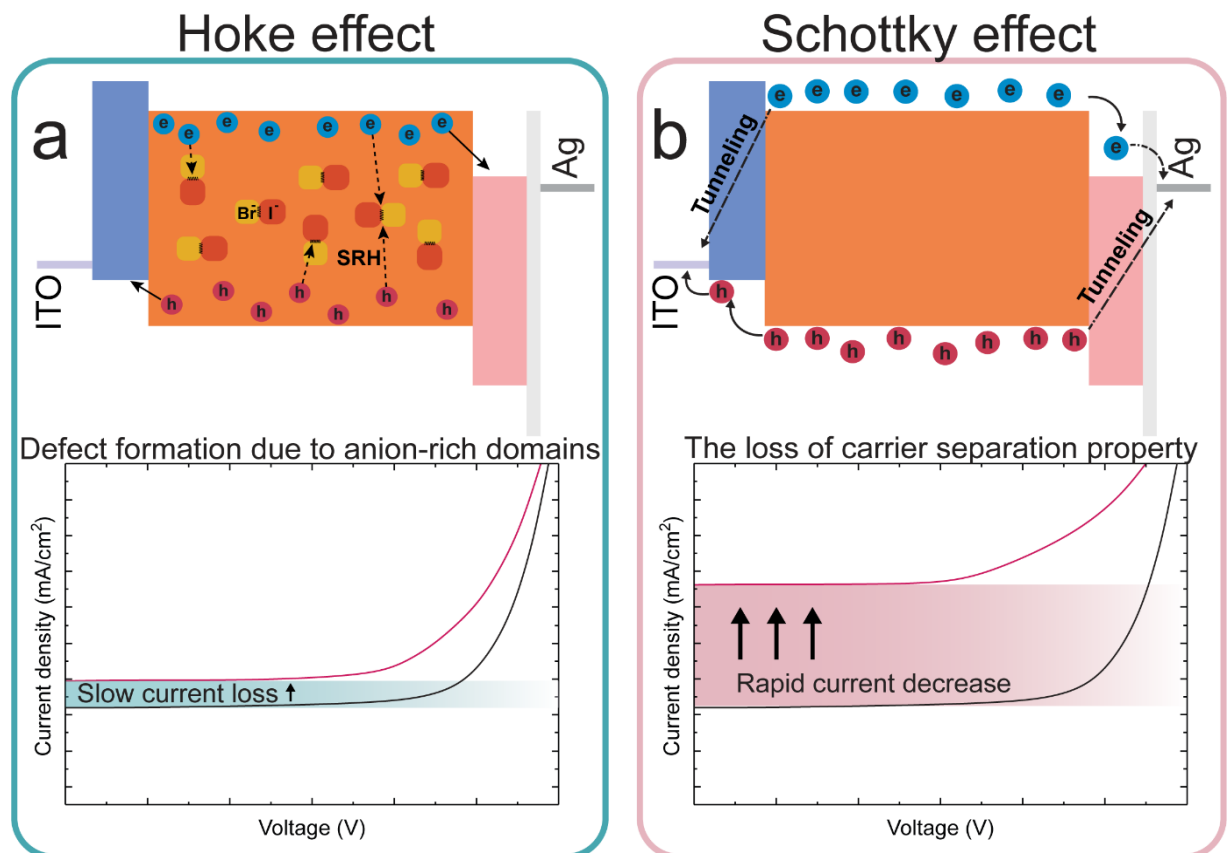

Figure 1. Hoke and Schottky effects in mixed halide perovskite SC. **a** Influence of Hoke effect on the performance of the mixed halide perovskite SC mainly determined by formation of defects and enhancing SRH recombination inside the perovskite layer, which results in slow current loss. **b** Schottky effect mainly downgrades the charge separation property of SC by effective thinning the TLs, leading to rapid current decrease.

The difference in these two degradation mechanisms lies in different parts of SC device. Hoke effect results in the defect formation inside the perovskite layer under sun illumination and formation of domains enriched with one type of halide (Figure 1a). These domains possess different lattice parameters, and the strain occurring between the host perovskite and halide domain results in consequential defect accumulation under sun illumination. The increasing number of defects results in improved recombination of charge carriers via defects with following decrease of the SC performance by gradual lowering of all main parameters: the open circuit voltage ($V_{OC}$), short-circuit current ($J_{SC}$), fill factor (FF) and power conversion efficiency (PCE). Schottky effect affects the charge separation property of SC device via modification of the device band structure due to formation of ion accumulation layers at the interfaces of perovskite with TLs. The blocking property and total charge carrier collection selectivity of transport layer deteriorate via introduction of image forces created by accumulation layers that lowers barrier making it effectively thin and thus downgrading the SC performance (Figure 1b). The loss of charge carrier selectivity also harms all parameters of SC as in case of Hoke effect. However, we assume that $J_{SC}$ is a principal figure of merit in both cases. $J_{SC}$ reflects both effects: charge-carrier selectivity and increase of defects in absorber layer. Its temporal behavior can be used to distinguish these two effects.

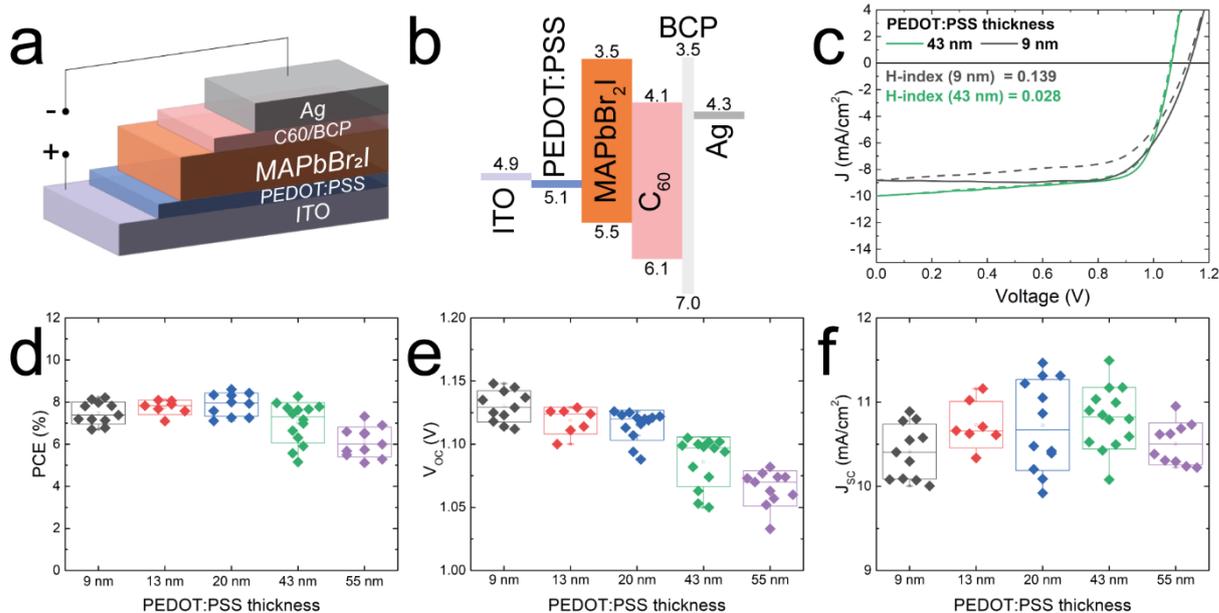

Figure 2. Device structure and SC performance. **a** Structure of p-i-n perovskite SC and **b** its band diagram. **c** JV curves of 9nm and 43nm PEDOT:PSS based devices, measured in forward (dash lines) and reverse (solid lines) directions. **d-f** Photovoltaic parameters of perovskite SC with different PEDOT:PSS thickness, FF statistics can be found in Figure S5.

Under applied voltage or sun illumination the ions inside the perovskite layer move towards the TLs according to their charge state ($MA^+$, $I^-$, $Br^-$) leading to several consequences: the first one is the formation of additional defects in perovskite crystalline structure, the second one is the formation of charged accumulation layers of ions at the perovskite/TL interfaces resulting in buildup of image forces promoting the tunneling of charge carriers through the TL. The defects inside the perovskite layer are formed due to ion movement and crystalline strain which is the result of formation and enlargement of halogen rich domains. The rate of defect formation inside the perovskite layer depends only on illumination intensity, regardless of the thickness of TL. Since charge carrier tunneling probability through the TL depends only on its thickness, it is possible to investigate this effect just by adjusting the TL thickness, thus opening the possibility to separate influences of Hoke and Schottky effects.

Having that we propose the following simple experiment, to distinguish and show the contributions of Hoke and Schottky effects on SC instability: we fabricated SC with the following TLs: ITO/PEDOT:PSS/MAPbBr$_2$I/C60/BCP/Ag (Figure 2a and 2b) with different PEDOT:PSS thicknesses (Figures S3, S4). We intentionally selected PEDOT:PSS due to the possibility of fine thickness tuning by spincoating parameters. The thicknesses of $C_{60}$ and BCP were fixed at 25nm and 8nm respectively as the most common thicknesses to achieve high SC performance[33,34].

Figure S6 shows typical JV characteristics for the devices based on different PEDOT:PSS thickness. The PCE values do not demonstrate significant changes regardless of the HTL thickness (Figure 2d), but they were achieved by gradual improvement of $J_{SC}$ and decrease of $V_{OC}$ with the increase of HTL thickness (Figure 2e and 2f). These changes reflect improvement in charge carrier separation property although the increased PEDOT:PSS thickness leads to strong recombination and thus losses in $V_{OC}$[35].

The best approach for description of the SC performance bases on the concepts of carrier collection probability and contact selectivity[36]. In terms of contact selectivity, the SC performance depends on TLs electronic properties: charge mobility, doping density, recombination losses, and layer width[37]. Usually for achievement of high PCE the TL thickness is set in 10-15 nm range[38,39] that grants high $J_{SC}$ value due to optimal thickness that allows to avoid the unnecessary losses and is sufficient to block the movement of opposite type of charge carriers to opposite electrode[40]. The carrier collection probability and contact selectivity are important parameters for the description of SC that directly affects the $J_{SC}$ value. The $J_{SC}$ value and its temporal behavior reflect all changes inside the SC regardless of their origins. Therefore, the variations in $J_{SC}$ can be used as a fine metric to track the changes in devices.

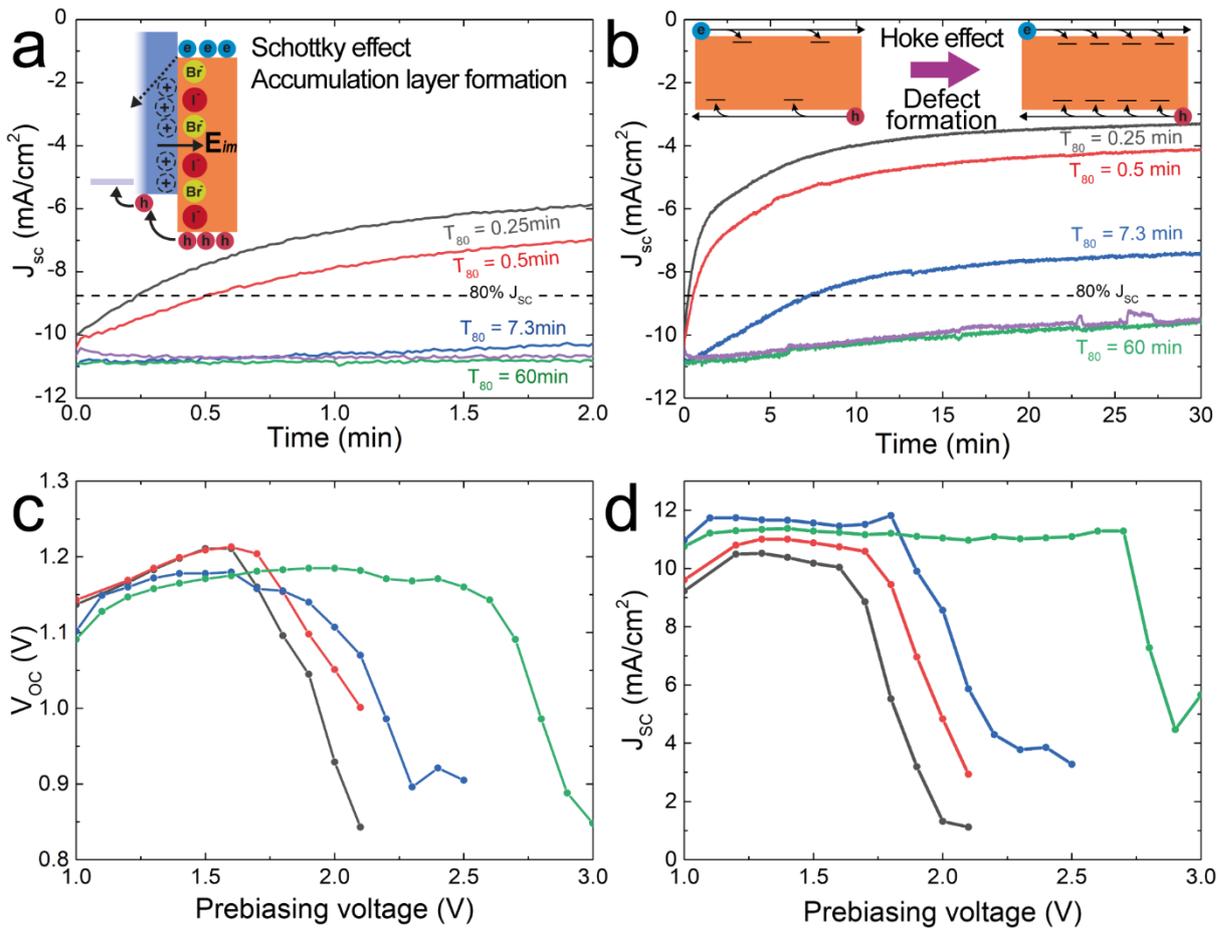

Figure 3. Sun soaking and voltage biasing. Time- and bias-dependent measurements of $J_{SC}$ and $V_{OC}$ of the devices with different thicknesses of PEDOT:PSS. $J_{SC}$ time laps under sun soaking at short **a** and long **b** times. At short times the rapid decrease of $J_{SC}$ corresponds to the manifestation of Schottky effect and slow $J_{SC}$ decrease at long sun exposure times corresponds to Hoke effect. **c** $V_{OC}$ stability against bias voltage, the thicker PEDOT:PSS allows to withstand stronger accumulation layers formed by prebias voltage before breakdown. **d** $J_{SC}$ dependence on prebias voltage demonstrating higher stability for thicker PEDOT:PSS layers preventing the charge tunneling through the transport layer owing to image forces created by accumulation layers.

Collection probability describes the probability of charge carriers to reach the electrodes and give a contribution in $J_{SC}$ value. Generally, $J_{SC}$ is affected by non-radiative (Shockley-Read-Hall, SRH) recombination inside the absorber material or TL[37], when photogenerated charges recombine in perovskite layer via defect states and do not contribute to overall $J_{SC}$. In our case the increase of HTL thickness results in improvement of $J_{SC}$ and preservation of PCE despite recombination losses in TL and overall $V_{OC}$ loss (Figure 2e and 2f) and in terms of described approaches we see that charge collection selectivity improves with the thicker PEDOT:PSS. The highest $J_{SC}$ value was obtained with the HTL thickness of 43nm.

Similarly, the hysteresis index (H-index) (Figure 2c) demonstrates improvement with thicker HTL. It is calculated using formula (1)[41], where rev and for stand for reverse and forward scan directions.

$$HI = \frac{PCE_{rev} - PCE_{for}}{PCE_{rev}} \quad (1)$$

Although the H-index in halide perovskites is a controversial parameter that can be used as figure of merit of SC performance but due to its dependence on voltage scan rate it rather to be used for evaluation of transient states of the device. Therefore, the use of identical scan rate of JV curve for the devices differing in HTL thickness can show how transient processes affect the device performance depending on device architecture. The thick HTL layer results in lower H-index pointing to the lower influence of ion migration on the overall device performance[41,42].

Along with H-index it is better to perform steady state measurements of SC like maximum power point tracking under continuous Sun exposure capable of demonstrating how fast the device performance reaches 80% of initial value ($T_{80}$ time). In our case we performed the $J_{SC}$ tracking under Sun illumination (Figure 3a and 3b). The time dependent $J_{SC}$ curves behavior demonstrated strong dependence on the HTL thickness, the curves have two distinct regions of $J_{SC}$ decrease. At short times of Sun exposure, the devices with thin HTL demonstrated a rapid decrease of $J_{SC}$ value before reaching the region of slow $J_{SC}$ decrease. The devices with thick HTL did not demonstrate this feature and devices with extremely thick HTL on the contrary demonstrated initial growth of $J_{SC}$ value (Figure 3a). Over a long time of Sun exposure all devices demonstrated similar $J_{SC}$ decrease rate.

Among the processes of ion migration and defect formation the first one is much faster than the second one and the rapid changes of $J_{SC}$ values at short times can be attributed to a manifestation of Schottky effect. In the devices with thin HTL Schottky effect manifests at short times of Sun exposure as rapid decrease of $J_{SC}$ and the devices with thick HTL do not demonstrate any rapid decrease or show even slight improvement of the $J_{SC}$ value (Figure 3a violet curve). While Hoke effect manifests as a slow decrease of $J_{SC}$ due to gradual formation of new defect states and thus enhancing the SRH recombination. The Schottky effect cause is the image forces arising due to the formation of the ion accumulation layers at TL/perovskite interfaces that result in effective thinning of TL allowing the tunneling of charge carriers (in our case it is electrons) to the electrode thus lowering carrier selectivity of the SC and $J_{SC}$ value. In devices with thin HTL it results in rapid decrease of $J_{SC}$ and in devices with extremely thick HTL it allows to slightly improve the $J_{SC}$ (Figure S7). JV curves of the samples before and after Sun soaking provided in Supporting Information (Figure S8).

The same situation occurs during voltage poling of the devices (Figure 3c and 3d). During voltage biasing ions pulled toward TL interfaces and form accumulation layer, but due to much higher poling voltage than $V_{OC}$ it is possible to form stronger layer and investigate the breakdown of the devices. In the case of thin HTL the devices (Figure 3c black and red curves) gradually demonstrated improvement of $V_{OC}$ and $J_{SC}$ indicating improvement of band alignment due to formation of pin structure. At certain voltage, which is labeled as $V_{cr}$, device breakdowns by rapid lowering of $V_{OC}$ and $J_{SC}$ value. The poling voltage that results in device breakdown depends on HTL thickness and increases with thicker HTL which corresponds to our concept model (Table S1).

The bias of the device leads to ion movement towards interfaces with TL with following accumulation layer formation. The charges of perovskite ions in these accumulation layers form image forces, which induce effective thinning of TL allowing tunneling of charge carriers through the TL layer, which, in general, manifests in the loss of charge carrier separation property and temporal degradation of $J_{SC}$. Performance decreasing of samples after different prebiases is shown in Supporting Information (Figure S9). After removing the exposure, the devices partially recovered their original performance, if maximum prebias was lower, than $V_{cr}$ (Figure S10).

For the verification of ion induced changes in charge collection selectivity we performed series of impedance measurements of devices with thin and thick HTL (9 and 43 nm) under different soaking and biasing conditions (Figure 4). During these measurements it is possible to distinguish between electronic and ionic contributions because of frequency dependent responses: ions are larger and contribute only at low frequencies. Large capacitance at low frequency is related to slow ionic polarization under electric field, while frequency increasing leads to capacitance lowering showing contribution of only electron subsystem[43]. Moreover, -Im(Z)(Re(Z)) dependence (Nyquist plot) allows constructing a model that incorporates ion diffusion and enables the study of its behavior under external stimuli, such as electrical prebias or Sun soaking[44]. Finally, capacitance-voltage dependencies show the result of charge accumulation under electric field, which is related to ion migration inside the perovskite film.

The CV measurement as well as $J_{SC}$ soaking were performed under different conditions: Sun soaking and voltage biasing. In both cases ions move towards electrodes providing reconfiguration of band structure. The behavior of capacitance-voltage (CV) characteristics measured under Sun exposure differ depending on the thickness of HTL (Figure S11). The main feature obtained from CV dependencies is the possibility to extract the built-in voltage ($V_{bi}$)[45] and in case of halide perovskite it is also possible to see how mobile ions affect the stability of electric characteristics of the device. In case of thin HTL the CV characteristic measured during continuous Sun exposure demonstrated significant changes unlike in case of thick HTL (Figure 4a and 4b). Mott-Schottky analysis demonstrated considerable shift of the linear part of the $C^{-2}$ dependence (Arrow in Figure 4a) responsible for the presence of depletion layer ($C_{dl}$), which is masked by the changes produced by mobile ions movement preventing correct determination of $V_{bi}$[46] for the thin PEDOT:PSS. In case of thick HTL linear part of the dependence stays almost intact, indicating the absence of aftereffects of mobile ions movement on the manifestation of depletion layer on $C^{-2}$ dependence. The $V_{bi}$ determined via linear approximation of $C_{dl}$ region gives value of 1.04 V (Figure 4b) and it grows with time under Sun soaking up to 1.19 V which corresponds to the data of $V_{oc}$ tracking (Figure 3c).

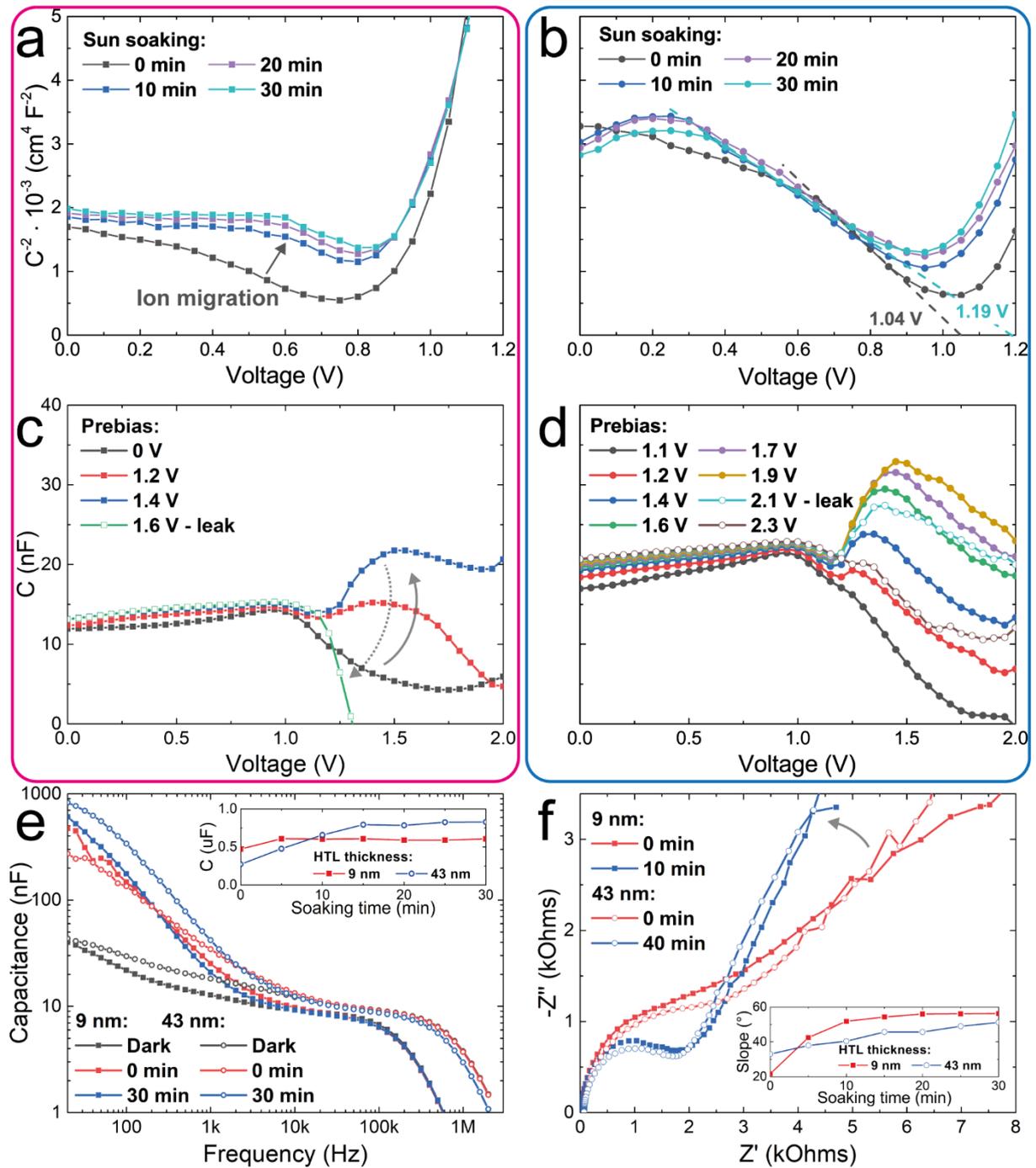

Figure 4. Capacitive characteristics of mixed halide perovskite SC with thin and thick HTL. **a** and **b** Capacitance-voltage dependencies measured during soaking under Sun exposure of the SC with thin and thick HTL (9 and 43 nm respectively). **c** and **d** Capacitance-voltage dependencies measured in the dark after voltage biasing of the SC with thin and thick HTL. **e** Cf dependencies of the SC with thin and thick HTL measured during the Sun exposure. The inset shows the dependence of the capacitance measured at 20Hz. **f** Impedance curves measured during the Sun exposure of SC with thin and thick HTL. The inset shows how the angle of low frequency arc changes over the Sun exposure.

CV characteristics measured in dark for the devices with thin or thick HTL demonstrate similar trends after the prebiasing at certain voltage (Figure 4c and 4d). The dependencies demonstrate an increase in the initial capacitance and the emergence of the second peak at higher voltage. The main difference is in the value of the prebias voltage that results in the breakdown of CV characteristic integrity. Both cases demonstrate the growth of capacitance at 0 V that corresponds to device charging owing to displaced ions produced during voltage prebias. Another similar and interesting feature is the emergence of the second peak at higher voltage that intensity growth with increasing of the prebias voltage. This peak arises due to reconfiguration of the device band structure due to formation of p-i-n structure allowing it to work as a light-emitting device which we demonstrated earlier[47]. In this case the peak position corresponds to voltage at which electrons are injected into the perovskite layer[48]. The main difference in the prebias voltage at which this peak disappears and the integrity of the CV characteristic is destroyed (open dot curves in Figure 4c and 4d). The breakdown of CV characteristics happens at 1.6 V (for thin HTL) and 2.0 V (for thick HTL) – these values are marked as "leak" on Figure 4c and 4d. In these points the loss of charge carrier selectivity for one type of charge carriers, in our case to electrons, occurs.

Cf characteristics measured under Sun soaking and short circuit conditions (Figure 4e) also demonstrated similar trends depending on the HTL thickness. In both cases low frequency component responsible for ion dynamics[49] demonstrates consequential growth during sun soaking but in case of thin HTL the final value is much lower than in case of thick HTL (500 nF and 900 nF respectively) (Figure 4e inset). Moreover, for the thick HTL the growth of low frequency capacitance demonstrates saturation behavior while in case of thin HTL it just pins to the certain value indicating rigid upper border such as inability to accumulate more charges due to leakage the lossy transport layers.

Impedance measurements performed during the Sun soaking of SC also demonstrated similar trends for thin and thick HTL (Figure 4f). In both cases impedance curves demonstrate Warburg-like feature, which manifests at low frequencies and represents as a line with certain angle[50]. This feature arises from the accumulation of ions on the perovskite interfaces with transport layers. Ideally, when the ion accumulation layer is formed it blocks all the charge transfer and line should be vertical due to the device capacitive charging[51], but since the devices demonstrate both ionic and electronic conduction the angle is much less than 90°. The time dynamics (inset on Figure 4f) of Warburg-like feature evolution under Sun exposure demonstrate fast (10 minutes) development to the final angle for devices with thin HTL while in devices with thick HTL it takes much longer time ( > 30 minutes). Which corresponds to the performance degradation measured during Sun exposure, where the JV curves demonstrated significant differences in the $J_{SC}$ value for thin HTL (45% loss of initial value) and thick HTL (only 14% loss) in 35 minutes.

It is important to note the specific behavior of Warburg feature during the Sun exposure – it increases until it reaches the final value. It is argued that alone Warburg feature does not necessarily point to the presence of ion migration but in our case, we see it tends to become vertical gradually increasing the angle along with the time of Sun exposure. Since these impedance measurements are generally treated from the transmission line point of view such vertical trend associates with termination of charge transfer from the perovskite layer to the electrode. The gradual increase of Warburg feature angle can be explained as consequential accumulation of ions at perovskite interfaces with transport layers along with reconfiguration in the ionic-electronic carrier system towards the higher low frequency impedance due to stronger accumulation layer and filled ionic reservoir.

Additional impedance measurements were conducted to investigate Warburg feature after high voltage prebias for the devices with thick PEDOT:PSS layer. The value of prebias voltage gradually increased from 1.1 V to 3.0 V. Samples were exposed with Sun during the entire process. It is shown that the slope is increasing with voltage (Figure S12), which corresponds to previous results and indicates ion accumulation at TL/perovskite interfaces. It is worth noting that after certain prebias value, close to and higher of $V_{CR}$ of thick PEDOT:PSS, Nyquist plot changes significantly (Figure S13), showing irreversible changes in devise structure and consequent degradation (Figure S10). This is a result of accumulated ions diffusion into transport layers which changes band structure and device composition[49,52].

The transition frequency value ($f_d$ = 1269Hz) derived from crossover between low and high frequency arcs allows to determine the chemical diffusion coefficient of migrating ion according to the perovskite layer thickness (d = 450nm) by simple formula $\omega_d = \frac{D_\mu}{d^2}$[49]. Obtained value of diffusion coefficient $D_\mu = 4.1 \cdot 10^{-7}$ cm$^2$/s is greater that values obtained for bromine ($D_\mu = 8.4 \cdot 10^{-10}$ cm$^2$/s)[53] or iodine ($D_\mu = 3 \cdot 10^{-9}$ cm$^2$/s)[53] ions. In our case this value rather reflects the average diffusion coefficient due to the presence of multiple migrating ions.

## Summary

In present work we demonstrated how effect of ionic migration on mixed halide perovskite solar cell performance can be effectively nullified by simple adjustment of transport layer thickness. We showed that formation of ion accumulation layers affects the charge carrier separation characteristics via manifestation of Schottky effect induced by the image forces produced by fast formation of ion accumulation layers during Sun exposure or voltage biasing. As a result, charge-carriers tunnel through the transport layer lowering carrier separation property and consequently decreasing the Jsc value. The degradation of SC performance can be slowed down by simple increase of transport layer thickness revealing the degradation associated with the defect formation under sun exposure, ion migration and Hoke effect. Nullification of the Schottky effect allows to immediate improvement of the mixed halide perovskite up to 240 times increasing the $T_{80}$ time from 15 seconds up to 60 minutes.

## Conflict of interests

There are no conflicts to declare.

## Data availability statement

All data which support this study are included in the published Article and its supplementary information (SI). DOI:

## Acknowledgements

The synthesis of perovskite was supported by the Russian Science Foundation (project № 21-73-20189-P), this work was supported by the Priority 2030 Federal Academic Leadership Program. The investigation particularly was carried out at the IRC for Nanotechnology of the Research Park of St. Petersburg State University within the framework of project No. №125021702335-5.

Supplementary information

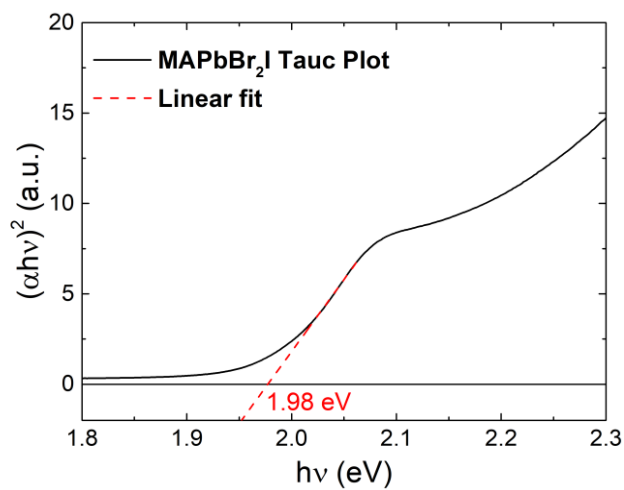

Figure S1. Tauc plot of MAPbBr$_2$I perovskite. Calculated E$_g$ is 1.98 eV.

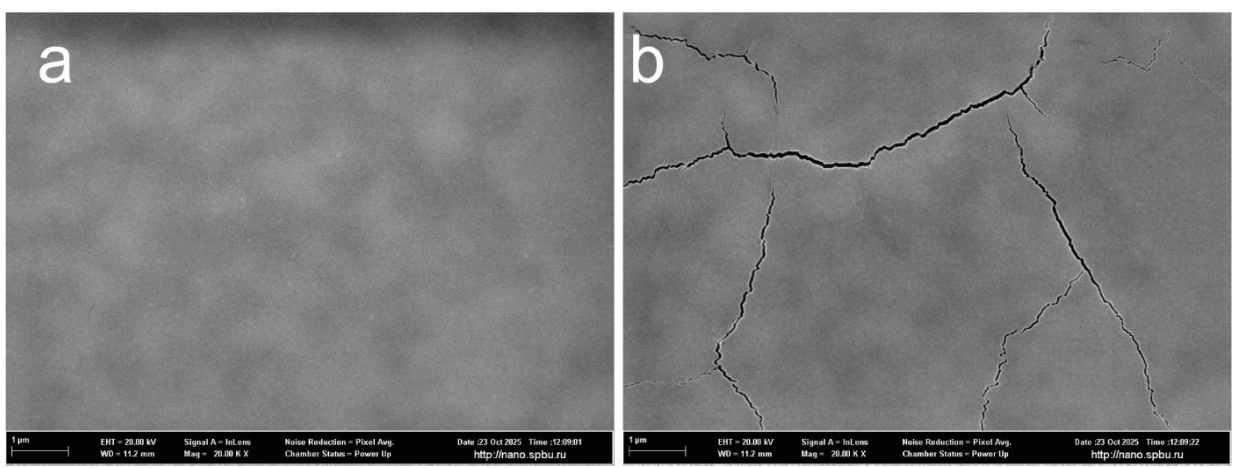

Figure S2. SEM images of the surface of MAPbBr$_2$I initial (a) and after short time of measurements (b). Cracks occur due to instability of the perovskite against high energy electrons.

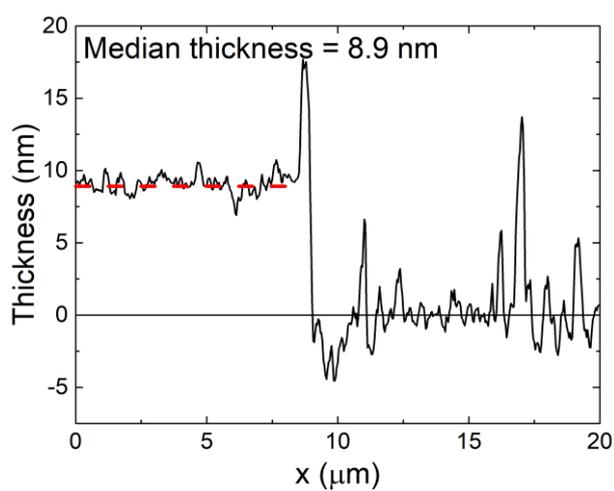

Figure S3. AFM profile of the thinnest PEDOT:PSS film for devices.

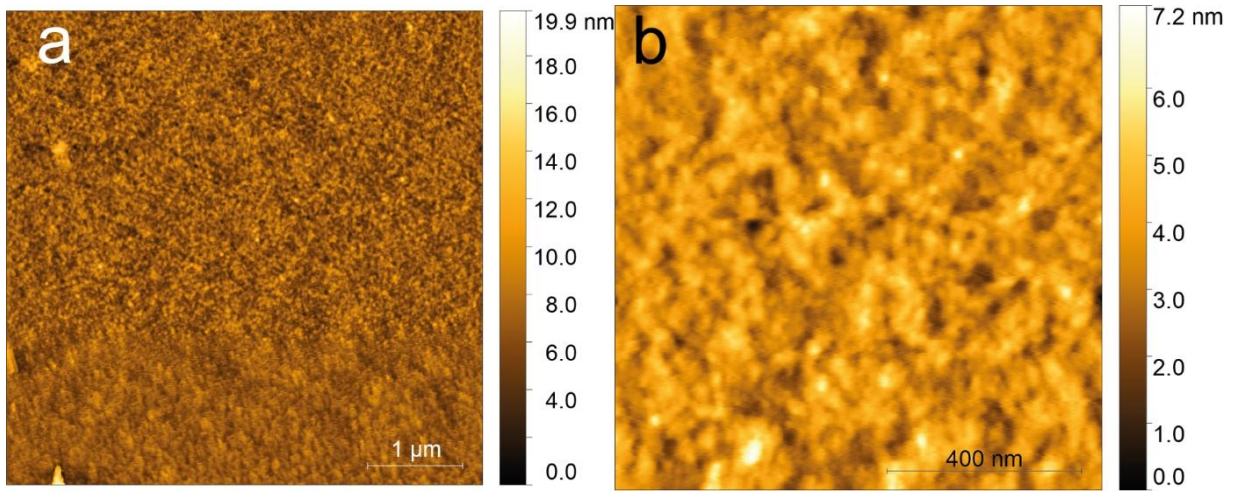

Figure S4. AFM surface images of the thinnest (9 nm) PEDOT:PSS film: RMS Roughness (Sq) for the area of 5x5 um is 1.37 nm (a), for the area of 1x1 um – 0.74 nm (b).

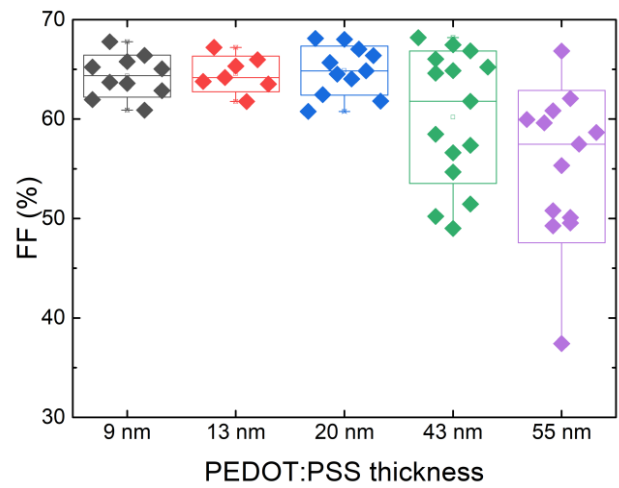

Figure S5. Statistics of FF for the devices with different PEDOT:PSS thickness.

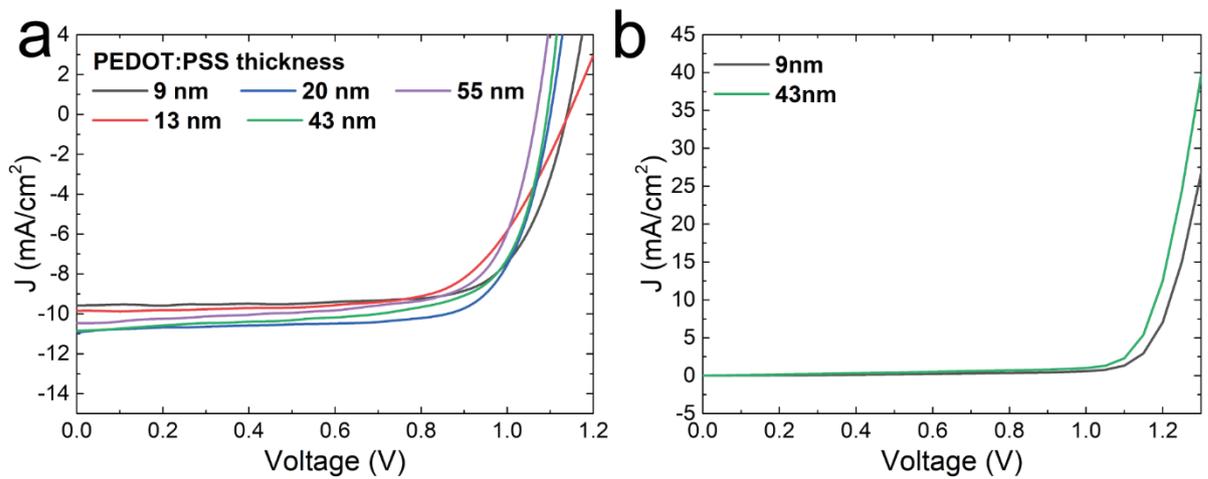

Figure S6. JV curves of devices with different PEDOT:PSS thickness (a). Dark JV curves for 9 and 43nm PEDOT:PSS thickness devices (b).

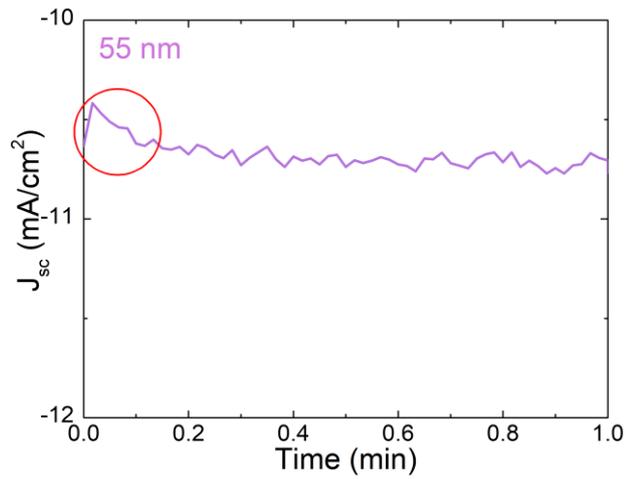

Figure S7. Slight improvement of Jsc for the device with extremely thick PEDOT:PSS layer (55 nm).

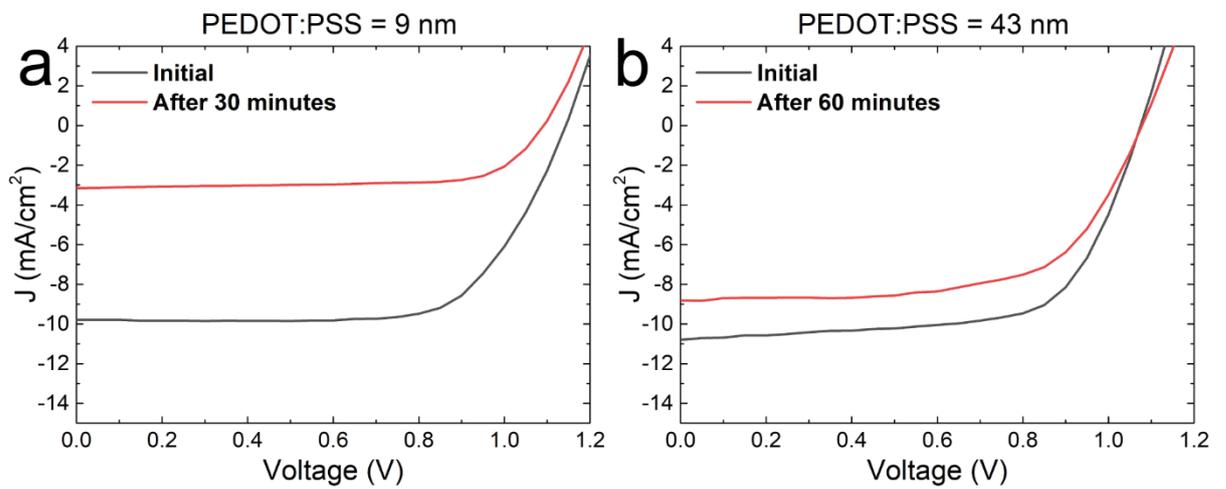

Figure S8. JV curves before and after Sun soaking for samples with thin (a) and thick (b) PEDOT:PSS.

Table S1. Critical values of voltage from $V_{oc}$ and $J_{sc}$ tracking.

| Thickness (nm) | $V_{cr}$ (V) | |
| --- | --- | --- |
| | from $V_{oc}$ tracking | from $J_{sc}$ tracking |
| 9 | 1.61 | 1.62 |
| 13 | 1.66 | 1.71 |
| 20 | 1.79 | 1.78 |
| 43 | 2.60 | 2.70 |

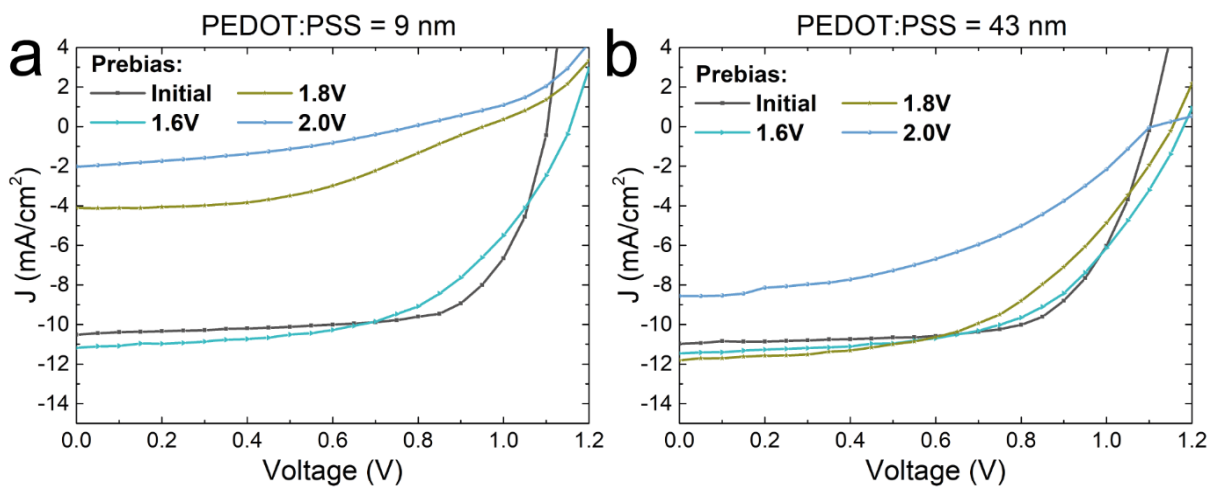

Figure S9. JV curves before and after electrical prebiases for samples with thin (a) and thick (b) PEDOT:PSS.

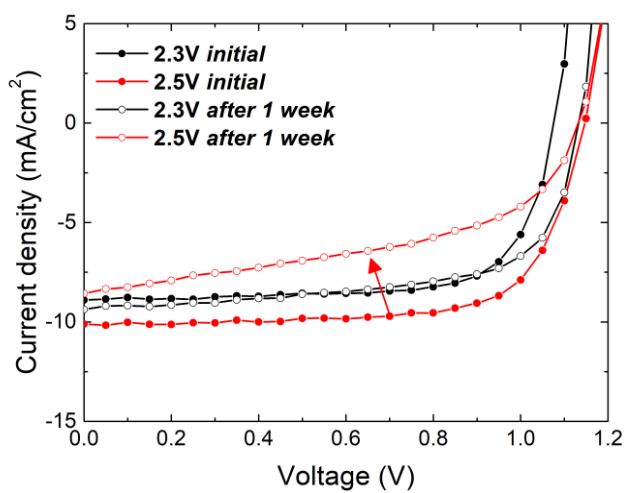

Figure S10. Degradation of the device with thick PEDOT:PSS layer (43 nm) after high prebias due to ion diffusion into transport layers (red curve). If prebias value is less than $V_{cr}$, device recover the original performance.

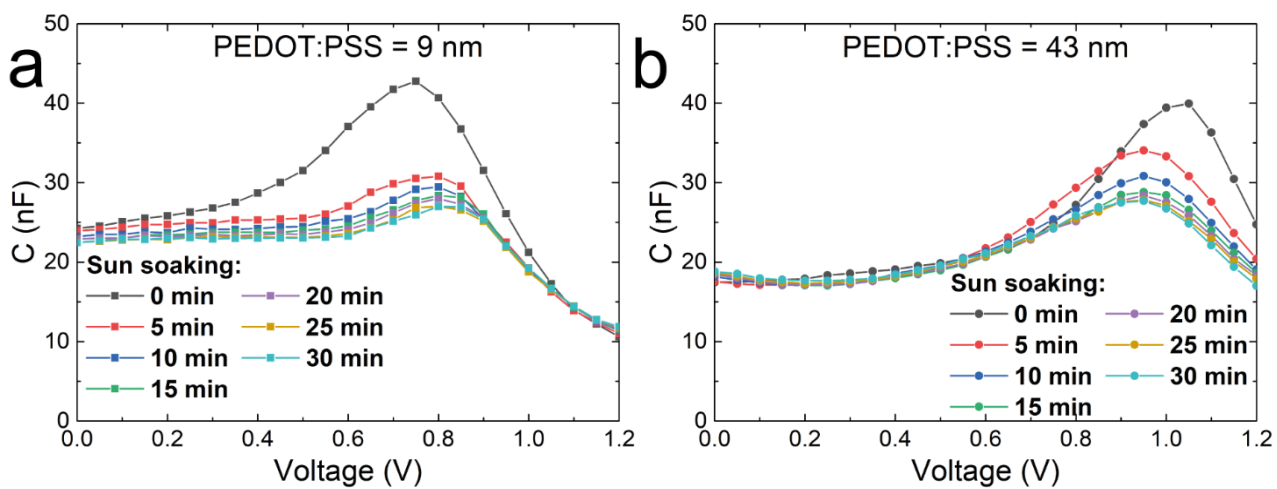

Figure S11. CV curves measured under Sun exposure for samples with thin (a) and thick (b) PEDOT:PSS.

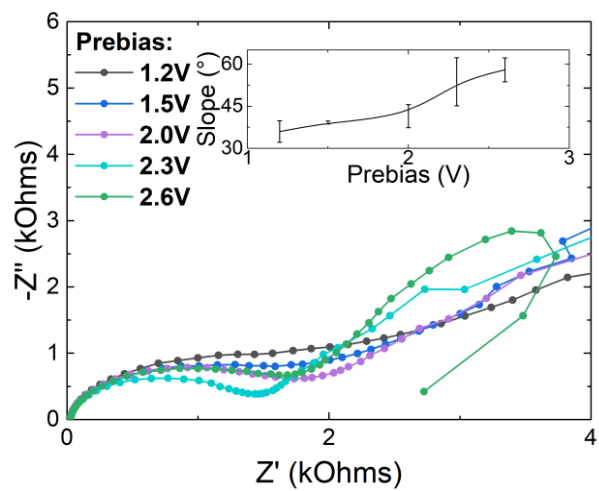

Figure S12. Nyquist plot of the sample after different prebiases. The slope of Warburg feature is increasing with prebias value (inset).

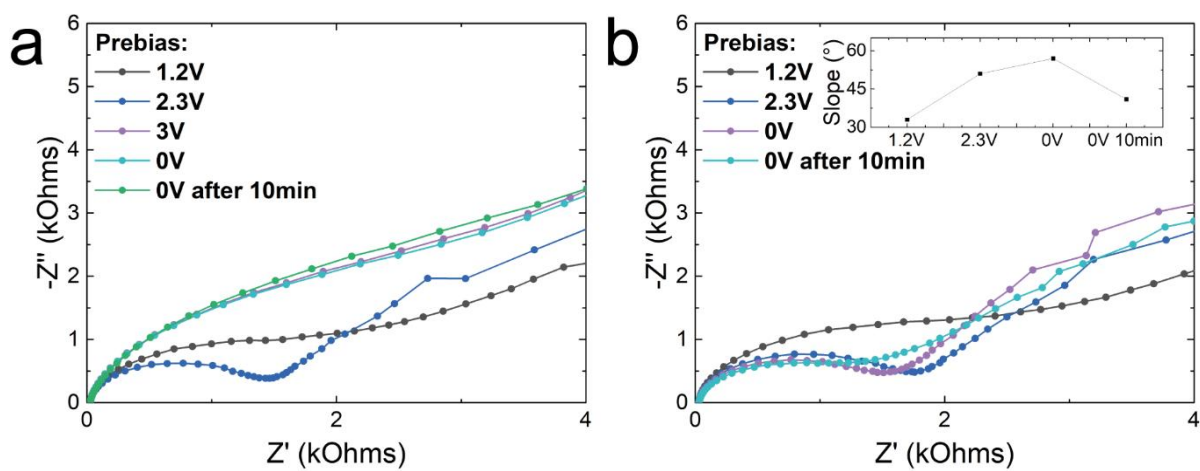

Figure S13. Nyquist plots of the sample after different prebiases and after measurement without prebias (0V) and after 10 minutes (0V after 10min). After high prebias (3V) sample does not return to its original shape, the changes in device are irreversible (a). After low prebias (less than $V_{cr}$) sample slowly returns to its original shape – the Warburg slope decreases (b).